\documentclass[aps,prl,reprint,superscriptaddress,floatfix]{revtex4-1}

\usepackage{graphicx}
\usepackage{siunitx}
\usepackage{hyperref}
\usepackage{color}

\usepackage[normalem]{ulem}

\begin{document}

\newcommand*\mycommand[1]{\texttt{\emph{#1}}}

\author{Kathrin Ganzhorn}
\email{kathrin.ganzhorn@wmi.badw.de}
\affiliation{Walther-Mei\ss ner-Institut, Bayerische Akademie der Wissenschaften, 85748 Garching, Germany}
\affiliation{Physik-Department, Technische Universit{\"a}t M{\"u}nchen, 85748 Garching, Germany}

\author{Tobias Wimmer}
\affiliation{Walther-Mei\ss ner-Institut, Bayerische Akademie der Wissenschaften, 85748 Garching, Germany}
\affiliation{Physik-Department, Technische Universit{\"a}t M{\"u}nchen, 85748 Garching, Germany}

\author{Joseph Barker}
\affiliation{Institute for Materials Research, Tohoku University, Sendai 980-8577, Japan}

\author{Gerrit E. W. Bauer}
\affiliation{Institute for Materials Research, Tohoku University, Sendai 980-8577, Japan}
\affiliation{WPI Advanced Institute for Materials Research, Tohoku University, Sendai 980-8577, Japan}
\affiliation{Zernike Institute for Advanced Materials, University of Groningen, 9747 AG Groningen, The Netherlands}

\author{Zhiyong Qiu}
\affiliation{WPI Advanced Institute for Materials Research, Tohoku University, Sendai 980-8577, Japan}

\author{Eiji Saitoh}
\affiliation{Institute for Materials Research, Tohoku University, Sendai 980-8577, Japan}
\affiliation{WPI Advanced Institute for Materials Research, Tohoku University, Sendai 980-8577, Japan}
\affiliation{CREST, Japan Science and Technology Agency, Tokyo 102-0076, Japan}
\affiliation{Advanced Science Research Center, Japan Atomic Energy Agency, Tokai 319-1195, Japan}

\author{Nynke Vlietstra}
\affiliation{Walther-Mei\ss ner-Institut, Bayerische Akademie der Wissenschaften, 85748 Garching, Germany}
\affiliation{Physik-Department, Technische Universit{\"a}t M{\"u}nchen, 85748 Garching, Germany}

\author{Stephan Gepr{\"a}gs}
\affiliation{Walther-Mei\ss ner-Institut, Bayerische Akademie der Wissenschaften, 85748 Garching, Germany}

\author{Rudolf Gross}
\affiliation{Walther-Mei\ss ner-Institut, Bayerische Akademie der Wissenschaften, 85748 Garching, Germany}
\affiliation{Physik-Department, Technische Universit{\"a}t M{\"u}nchen, 85748 Garching, Germany}
\affiliation{Nanosystems Initiative Munich, 80799 Munich, Germany}

\author{Hans Huebl}
\affiliation{Walther-Mei\ss ner-Institut, Bayerische Akademie der Wissenschaften, 85748 Garching, Germany}
\affiliation{Physik-Department, Technische Universit{\"a}t M{\"u}nchen, 85748 Garching, Germany}
\affiliation{Nanosystems Initiative Munich, 80799 Munich, Germany}

\author{Sebastian T.B. Goennenwein}
\affiliation{Walther-Mei\ss ner-Institut, Bayerische Akademie der Wissenschaften, 85748 Garching, Germany}
\affiliation{Physik-Department, Technische Universit{\"a}t M{\"u}nchen, 85748 Garching, Germany}
\affiliation{Nanosystems Initiative Munich, 80799 Munich, Germany}
\affiliation{Institut f\"ur Festk\"orper- und Materialphysik, Technische Universit\"at Dresden, 01062 Dresden, Germany}
\affiliation{Center for Transport and Devices of Emergent Materials, Technische Universit\"at Dresden, 01062 Dresden, Germany}

\title{Non-local magnon transport in the compensated ferrimagnet GdIG}

\begin{abstract}
We study the diffusive transport of magnons through the compensated ferrimagnetic insulator Gd$_3$Fe$_5$O$_{12}$ (GdIG). The magnons are injected and detected electrically in a non-local measurement configuration via two parallel Pt strips deposited on top of the ferrimagnet. GdIG exhibits a rich magnon spectrum, with several thermally populated magnon bands at room temperature.
We observe a strong temperature and field dependence of the non-local voltage in the detector strip. Just below the magnetization compensation temperature we find that the increasing magnetic field causes an unexpected enhancement of the non-local signal. A comparison with GdIG spin wave spectra obtained from atomistic modeling indicates that the thermal magnon population is important for understanding the non-local voltage signal.

\end{abstract}

\maketitle

Several experimental and theoretical investigations of diffusive magnon transport through the ferrimagnetic insulator yttrium iron garnet (Y$_{3}$Fe$_5$O$_{12}$, YIG) have recently been conducted, using two electrically isolated Pt strips for electrical injection and detection \cite{CornelissenNat2015, Goennenwein2015, ZhangPRL, CornelissenPRB94, VelezMMR, Li_MMR_trilayer, Wu_MMR}. Upon running a dc charge current through a Pt injector strip, an electron spin accumulation is generated at the YIG\textbar Pt interface via the spin Hall effect (SHE) \cite{Dyakonov, Hirsch}. This spin accumulation induces a non-equilibrium magnon accumulation in the magnet beneath the injector which diffuses away from the injector and is detected in a second Pt strip via the inverse spin Hall effect (ISHE). This so-called non-local magnon mediated magnetoresistance (MMR) has been studied as a function of magnetic field \cite{CornelissenField} and temperature in YIG/Pt bilayers \cite{Goennenwein2015, CornelissenPRB94.180402}, giving insight into the length scales associated with magnon transport. However, the microscopic mechanisms responsible for the MMR are still under investigation. One key question is which magnons or magnon bands provide the most significant contributions to diffusive spin transport and the non-local voltage signal induced in the detector strip. Furthermore, the MMR has only been measured in collinear magnetic systems such as YIG or NiFe$_2$O$_4$ \cite{ShanNFO} so far. The influence of non-collinear magnetic configurations, such as a canted ferrimagnet, has not yet been investigated. 

Here, we study the non-local magnon transport in the compensated ferrimagnet gadolinium iron garnet (Gd$_{3}$Fe$_5$O$_{12}$, GdIG), where the non-magnetic Y$^{3+}$ in YIG is substituted with magnetic Gd$^{3+}$. GdIG therefore consists of three magnetic sublattices: two antiferromagnetically coupled Fe$^{3+}$ sublattices (FeA and FeD), and a Gd sublattice which is weakly antiferromagnetically coupled to the FeD moments \cite{Dionne:Magnetic:Oxides:Book:2009, GanzhornPRB}. GdIG exhibits a magnetization compensation temperature $T_\mathrm{comp}$, where the remanent magnetization exactly vanishes due to the different temperature dependencies of the antiferromagnetically coupled net Fe and Gd sublattice magnetizations \cite{Dionne:Magnetic:Oxides:Book:2009}. 
Far away from the compensation temperature, GdIG is a collinear ferrimagnet and the sublattice magnetizations are all aligned (anti)parallel to the external magnetic field. Close to $T_\mathrm{comp}$, a canted phase can be induced with magnetic field magnitudes accessible in experiments, where the magnetic moments on the different sublattices are no longer collinear with the field, or one another \cite{Dionne:Magnetic:Oxides:Book:2009, Bernasconi, Clark, GanzhornPRB}. The corresponding magnetic phase diagram can be found in Ref.~\cite{GanzhornPRB}. Recently it was shown that the spin Hall magnetoresistance effect \cite{AltiPRB, ChenSMR, NakayamaSMR, SMR:Vlietstra:PRB:2013} (SMR) can electrically probe the interface magnetization texture, since the SMR is sensitive to the orientation of the individual sublattice magnetic moments relative to the polarization of the spin accumulation in the adjacent Pt \cite{GanzhornPRB, AqeelCSO}. Non-local magnon transport or spin diffusion has not yet been studied in a non-collinear magnetic structure and will be presented in this Article. 
Aside from the magnetic compensation point, GdIG also differs from YIG in that the spin wave spectrum is richer in the low THz regime. Several additional modes are present due to the magnetic Gd sublattice.  Local spin Seebeck effect (SSE) measurements in GdIG showed strong evidence that two spin wave modes with opposite polarization dominate the thermal and dynamical behaviour of the spin system, resulting in a sign change of the SSE voltage \cite{SSEGdIG, CramerSSE}. GdIG is therefore a good material to investigate the influence of different magnon modes and their polarization on spin diffusion. Our experimental results suggest that the complex magnon spectrum of GdIG indeed qualitatively impacts the field and temperature dependence of the MMR.

\begin{figure}[!ht]
\includegraphics[width=\columnwidth]{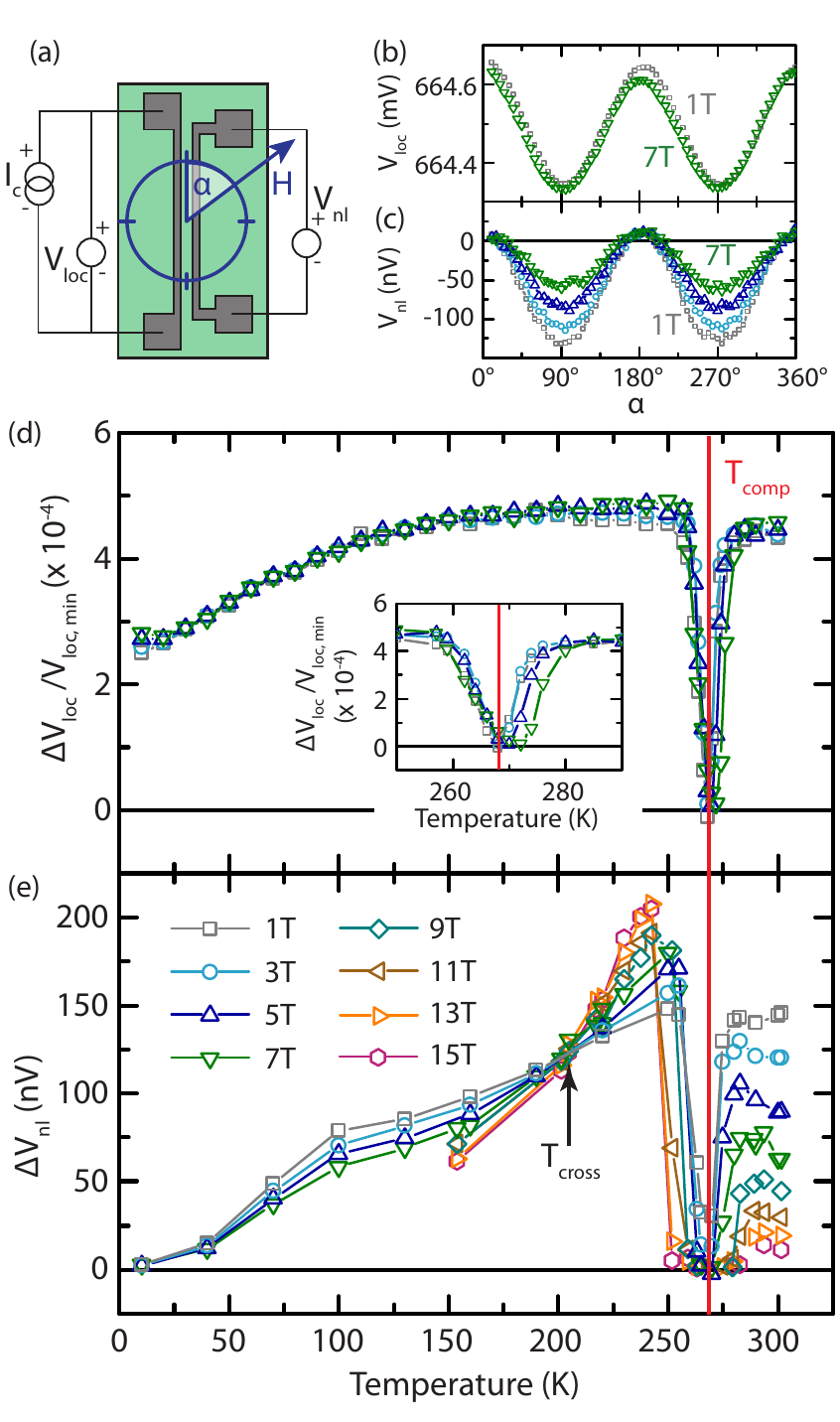}%
\caption{(a) Schematic representation of the GdIG(green)/Pt (grey) nanostructures consisting of two parallel Pt strips of width $w$ and edge-to-edge separation $d$. (b) Local voltage $V_\mathrm{loc}$ as a function of the angle $\alpha$ between $\textbf{I}_c$ and $\textbf{H}$ at \SI{300}{\kelvin}. Due to the weak field dependence of the local signal only experimental data taken at $\mu_0H=\SI{1}{\tesla}$ and \SI{7}{\tesla} are shown. (c) Non-local voltage $V_\mathrm{nl}$ as a function of $\alpha$ at \SI{300}{\kelvin} for $\mu_0H=1,3,5$ and \SI{7}{\tesla}. (d) Temperature dependent SMR amplitude $\left(V_\mathrm{loc}(\SI{0}{\degree}) - V_\mathrm{loc}(\SI{90}{\degree})\right)/ V_\mathrm{loc}(\SI{90}{\degree})$ extracted from the local voltage amplitude for different magnetic fields up to \SI{7}{\tesla}. The inset is a close-up of the SMR amplitude around the compensation temperature $T_\mathrm{comp}=\SI{268}{\kelvin}$ (red vertical line). (e) Temperature dependence of the MMR amplitude $\Delta V_\mathrm{nl}=V_\mathrm{nl}(\SI{0}{\degree}) - V_\mathrm{nl}(\SI{90}{\degree})$ obtained from ADMR measurements for magnetic fields up to \SI{15}{\tesla}.}%
\label{SMR_MMR}%
\end{figure}

The investigated sample is a \SI{2.6}{\micro\meter} thick GdIG film grown on top of a (111)-oriented Gd$_3$Ga$_5$O$_{12}$ substrate via liquid phase epitaxy (LPE). 
After cleaning in Piranha solution and annealing at \SI{500}{\celsius} in an oxygen atmosphere of \SI{75}{\micro\bar} for 40 min, a \SI{10}{\nano\meter} thick Pt film was deposited onto the GdIG via electron beam evaporation (see Refs.~\citenum{Goennenwein2015} and \citenum{PuetterSMR} for details). Two Pt strips with edge-to-edge separation $d=\SI{200}{\nano\meter}$ and strip width $w=\SI{500}{\nano\meter}$ (Fig.~\ref{SMR_MMR} (a)) were defined by electron beam lithography followed by Ar ion etching.
The sample was mounted in the variable temperature insert of a superconducting magnet cryostat allowing  dc transport measurements at temperatures between $\SI{5}{\kelvin} < T < \SI{300}{\kelvin}$ and magnetic fields up to $\SI{15}{\tesla}$.
For the electric transport measurements a charge current $I_c=\SI{100}{\micro\ampere}$ (corresponding to a current density of $2\times 10^{10}\si{\ampere\per\square\meter}$) was applied to the left strip using a Keithley 2400 sourcemeter. The local ($V_\mathrm{loc}$) and non-local voltage drop ($V_\mathrm{nl}$) in the injector and detector strip were simultaneously measured with two Keithley 2182 nanovoltmeters (see Fig.~\ref{SMR_MMR} (a)). To eliminate thermal signals due to Joule heating in the local strip (e.g. spin Seebeck effect \cite{Schreier_iSSE}) and to increase the signal-to-noise ratio, a current switching method was used, as described in Ref.~\cite{Goennenwein2015}. To obtain the SMR and MMR amplitude, angle dependent magnetoresistance (ADMR) measurements were carried out by rotating the external magnetic field of constant magnitude in the thin film plane, while recording $V_\mathrm{loc}$ and $V_\mathrm{nl}$. 

The local response $V_\mathrm{loc}$ measured at \SI{300}{\kelvin} as a function of the angle $\alpha$ between $\textbf{I}_c$ and $\textbf{H}$ is shown in Fig.~\ref{SMR_MMR} (b) for $\mu_0H=\SI{1}{\tesla}$ (grey) and \SI{7}{\tesla} (green). From these ADMR measurements, the SMR amplitude defined as $\Delta V_\mathrm{loc}/V_\mathrm{loc, min}=\left(V_\mathrm{loc}(\SI{0}{\degree}) - V_\mathrm{loc}(\SI{90}{\degree})\right)/ V_\mathrm{loc}(\SI{90}{\degree})$ was extracted and plotted as a function of temperature for magnetic field strengths $\mu_0H=1,3,5$ and \SI{7}{\tesla} in Fig.~\ref{SMR_MMR} (d). Cooling to lower temperatures, the SMR amplitude decreases by a factor 2 compared to room temperature, consistent with measurements in YIG/Pt \cite{MeyerSMRAPL} and (In, Y) doped GdIG/Pt bilayers \cite{GanzhornPRB}. In a narrow temperature range around the compensation temperature ($T_\mathrm{comp}=\SI{268}{\kelvin}$ determined via SQUID magnetometry), the SMR decreases to zero. A similar behaviour has been observed in (In, Y) doped GdIG \cite{GanzhornPRB} and was attributed to the formation of the canting phase in GdIG close to the compensation temperature, where the sublattice magnetizations are no longer collinear. In Ref.~\cite{GanzhornPRB} a sign change of the SMR was observed and interpreted as a perpendicular alignment of the sublattice moments with respect to the external field. The absence of such a sign change in the present data may be attributed to magnetic domain formation for temperatures close to $T_\mathrm{comp}$, such that on average the SMR modulation vanishes. However the broadening of the dip in the SMR vs.\@ temperature curves with increasing magnetic field (see inset of Fig.~\ref{SMR_MMR} (d)) is in good agreement with the observations in Ref.~\cite{GanzhornPRB} and is expected for the magnetic canting phase in GdIG \cite{Bernasconi}. As reported in Ref.~\cite{GanzhornPRB}, in the collinear phase the SMR in Fig.~\ref{SMR_MMR} (d) hardly depends on the applied magnetic field. This is consistent with the established SMR model \cite{ChenSMR, SMR:review:Chen:JPCM:2016}, where only the orientation of the sublattice moments is relevant \cite{GanzhornPRB} \footnote{The orientation of the sublattice moments is crucial for the SMR in non-collinear magnets. In the collinear phase, all magnetic moments are aligned along the same axis. Describing the SMR in terms of either the total or the sublattice magnetization is then equivalent.}.

The angular dependence of the non-local response $V_\mathrm{nl}$ at \SI{300}{\kelvin} is shown in Fig.~\ref{SMR_MMR} (c) for different magnetic field strengths. The non-local voltage signature has the same dependence on the external magnetic field orientation as in previous MMR measurements in YIG/Pt nanostructures for the same wiring scheme \cite{Goennenwein2015}. The MMR effect amplitude extracted as $\Delta V_\mathrm{nl}= V_\mathrm{nl}(\SI{0}{\degree})-V_\mathrm{nl}(\SI{90}{\degree}) = -V_\mathrm{nl}(\SI{90}{\degree}) > 0$ is plotted as a function of temperature in Fig.~\ref{SMR_MMR} (e) for fields up to \SI{7}{\tesla}. In contrast to the local SMR, where no substantial field dependence is observed in the collinear phase, the non-local MMR signal displays a more complex field and temperature dependence. We have therefore taken additional non-local data in fields up to \SI{15}{\tesla} for temperatures between \SI{150}{\kelvin} and \SI{300}{\kelvin}, as shown in Fig.~\ref{SMR_MMR} (e). We first focus on the data taken at \SI{1}{\tesla} (grey open squares in Fig.~\ref{SMR_MMR} (e)). At low temperatures $\Delta V_\mathrm{nl}$ vanishes, similar to recent observations in YIG/Pt \cite{Goennenwein2015} where a $T^{3/2}$ dependence of $\Delta V_\mathrm{nl}$ was observed. This temperature dependence in YIG is in agreement with theoretical expectations based on the magnon density of states and distribution function \cite{ZhangPRL, ZhangPRB, CornelissenPRB94}, i.e.~general properties of the magnonic system. At low temperatures, we therefore expect GdIG to behave very similar to YIG. 

With increasing temperature, the non-local signal measured at \SI{1}{\tesla} increases up to \SI{150}{\nano\volt} just below the compensation temperature. The MMR amplitude then drops below the experimental resolution of about $\SI{5}{\nano\volt}$ \cite{Goennenwein2015} at $T_\mathrm{comp}$ and recovers a finite value above $T_\mathrm{comp}$, similar to what is observed in the local SMR (see Fig.~\ref{SMR_MMR} (d)). We  attribute the vanishing MMR signal at $T_\mathrm{comp}$ to the change of the magnetic structure of GdIG into the canted phase. A vanishing non-local signal in the canted phase suggests two possible scenarios, either (i) the magnon injection/detection becomes inefficient due to the non-collinear alignment of the magnetic sublattice moments and/or (ii) the damping close to the compensation point is enhanced \cite{CalhounPR} leading to shorter magnon lifetimes and thus shorter diffusion lengths. Additional magnon scattering effects may arise due to magnetic domain formation in the canting phase as discussed above, which further suppress magnon transport.

\begin{figure}[!h]%
\includegraphics[width=\columnwidth]{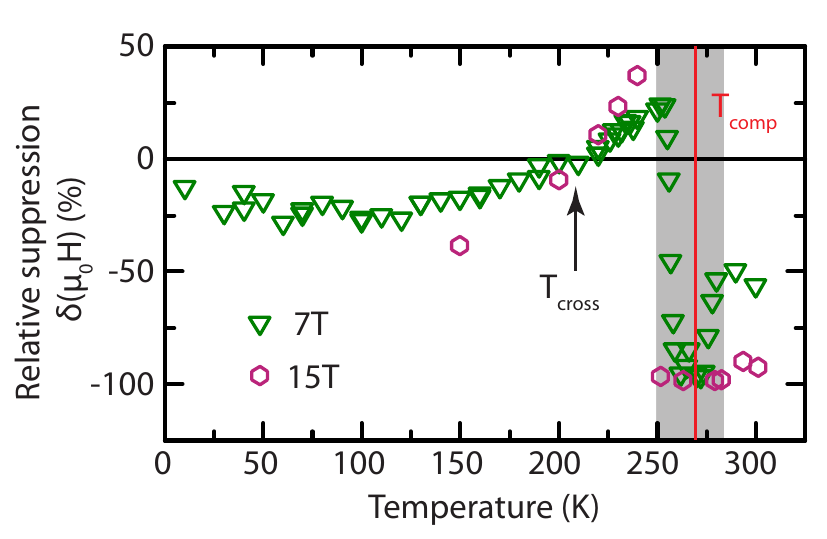}%
\caption{Relative magnetic field dependence $\delta(\SI{7}{\tesla})$ (green triangles) and $\delta(\SI{15}{\tesla})$ (purple hexagons) of the MMR effect defined by Eq.~(\ref{eq:1}). A negative (positive) value corresponds to a suppression (enhancement) with increasing field. The grey shaded region indicates the temperature range around $T_\mathrm{comp}$ in which the MMR drops to values close to or even below the noise level, impeding a reliable quantification of $\delta$.}%
\label{relsup}%
\end{figure}

We now turn to the magnetic field dependence of the MMR effect. The relative field dependence $\delta(\mu_0 H)$ of the MMR amplitude $\Delta V_\mathrm{nl}$ is defined as 

\begin{equation}
\delta(\mu_0 H) = \frac{\Delta V_\mathrm{nl}(\mu_0 H) - \Delta V_\mathrm{nl}(\SI{1}{\tesla})}{\Delta V_\mathrm{nl}(\SI{1}{\tesla})}
\label{eq:1}
\end{equation}
and plotted as a function of temperature in Fig.~\ref{relsup} for $\mu_0 H=\SI{7}{\tesla}$ (green triangles). A negative value corresponds to a suppression of the non-local voltage, while a positive value represents an increase of the MMR with increasing magnetic field. The field dependence is calculated relative to the measurements at \SI{1}{\tesla}, which is the lowest magnetic field studied here. We distinguish three temperature regimes: (i) for $0 < T < T_\mathrm{cross}\approx \SI{210}{\kelvin}$, the MMR amplitude is suppressed with applied magnetic field by about $\delta (\SI{7}{\tesla})=25\%$, (ii) in the range $T_\mathrm{cross} < T < T_\mathrm{comp}$ an enhancement with field is observed, while (iii) for $T_\mathrm{comp} < T $ the MMR effect is suppressed by up to $\delta (\SI{7}{\tesla})=50 \%$. 
The MMR suppression in the canted phase (grey shaded area in Fig.~\ref{relsup}) appears to be very large. However, since the MMR amplitude is close to the noise level, the relative field effect is prone to large errors. 
Additional measurements of $\delta(\mu_0 H)$ up to $\mu_0 H=\SI{15}{\tesla}$ between $T=\SI{150}{\kelvin}$ and \SI{300}{\kelvin} (purple hexagons in Fig.~\ref{relsup}) qualitatively confirm the data taken at \SI{7}{\tesla}, displaying a more pronounced suppression and enhancement of the MMR. In particular, the non-local signal can be almost completely suppressed above $T_\mathrm{comp}$ at \SI{15}{\tesla}.

A magnetic field induced reduction of the MMR at room temperature has also been observed in YIG/Pt and was attributed to a field-dependent magnon diffusion length and a reduced magnon injection efficiency \cite{CornelissenField}. Further temperature dependent MMR measurements on YIG/Pt (not shown here) confirm a field suppression of up to $\delta (\SI{7}{\tesla})\approx -25\%$ in the entire investigated temperature range from $\SI{50}{\kelvin}$ to $\SI{300}{\kelvin}$. While the low temperature field suppression of the MMR in GdIG/Pt is similar to that of YIG/Pt, the field dependence above $T_\mathrm{comp}$ is much stronger. Furthermore, the enhancement of the MMR with magnetic field in the temperature range $\SI{200}{\kelvin}$ to $\SI{250}{\kelvin}$ as shown in Fig.~\ref{SMR_MMR} and \ref{relsup} is also not observed in YIG/Pt.

In the following we discuss our experimental observations in terms of the magnon injection efficiency which depends on the available magnon states in the ferrimagnet \cite{ZhangPRB} and is thus related to the thermal magnon population. In YIG, only the lowest energy ferromagnetic-like  mode is thermally populated within the studied temperature range \cite{Harris_spinwave, BarkerYIG} (only near room temperature the high energy exchange mode starts to become populated and is expected to affect the SSE in YIG \cite{BarkerYIG}). An external magnetic field shifts the ferromagnetic mode to higher frequencies as the Zeeman gap opens, thereby freezing out the lowest energy magnons. In this picture, the magnetic field suppression observed in SSE experiments in YIG/Pt was interpreted in terms of the importance of low energy magnons at elevated temperatures \cite{KikkawaSSE, Boona_Heremans}. The suppression of the  detected MMR signal in YIG when increasing the external magnetic field at a fixed injector-detector separation \cite{CornelissenField} is also consistent with a freeze out of magnons from the ferromagnetic mode.

\begin{figure}[!ht]%
\includegraphics[width=\columnwidth]{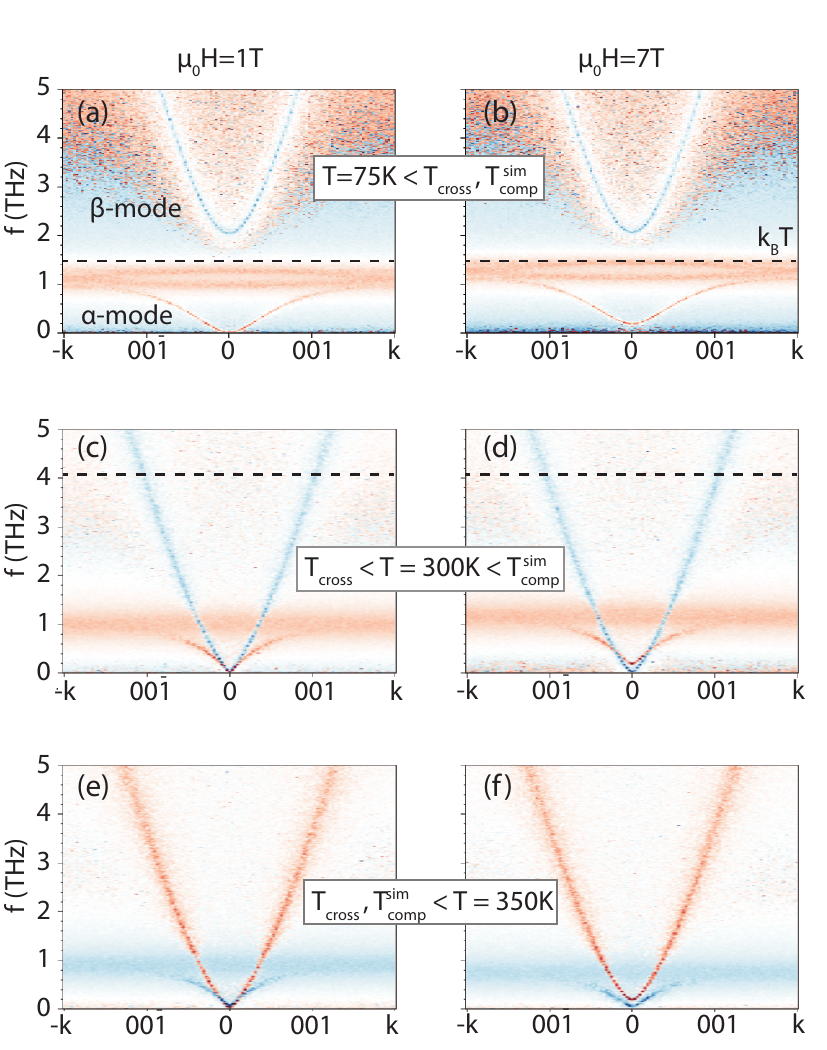}%
\caption{Magnon spectrum of GdIG obtained from atomistic simulations, calculated (a)-(b) at $T=\SI{75}{\kelvin}$, (c)-(d) just below $T_\mathrm{comp}^\mathrm{sim}=\SI{310}{\kelvin}$ and (e)-(f) above the compensation point. The red and blue color indicates counter-clockwise and clockwise precession directions relative to the external magnetic field direction, i.e. positive and negative polarization, respectively. The spectra are calculated for $\mu_0 H=\SI{1}{\tesla}$ and \SI{7}{\tesla} revealing the effect of the Zeeman shift of the red mode. The black dashed line corresponds to the thermal energy $k_\mathrm{B}T$.}%
\label{magnon_modes}%
\end{figure}

To better understand the observed complex field and temperature dependence of the MMR in GdIG we therefore study the characteristic changes in the GdIG spin wave spectrum as a function of temperature and magnetic field using atomic spin dynamics simulations. The calculations are based on a classical Heisenberg model where the Landau-Lifshitz-Gilbert equation is used to solve the spin dynamics and a Langevin thermostat introduces temperature. Details of the model including the parameters used for GdIG can be found in Refs.~\cite{SSEGdIG,GanzhornPRB}. The spin wave spectrum is calculated from the space-time Fourier transform of the spin fluctuations. The spin wave modes in ferrimagnets have a polarization which describes the sense of rotation with respect to an applied field~\cite{BarkerYIG}, i.e.~counterclockwise/clockwise, which is encoded by a red/blue coloring in the figures. We label the modes $\alpha$ - the lowest frequency dispersive mode at $\SI{75}{\kelvin}$ and $\beta$ - the parabolic optical mode with opposite polarization (see Fig.~\ref{magnon_modes} (a)). The modes move in frequency space with temperature and field and change polarization across the compensation point at $T_\mathrm{comp}^\mathrm{sim}=\SI{310}{\kelvin}$, but we retain the same $\alpha$, $\beta$ designations throughout. Note that $T_\mathrm{comp}^\mathrm{sim}=\SI{310}{\kelvin}$ is higher than the value $T_\mathrm{comp}=\SI{268}{\kelvin}$ experimentally observed in our sample. The flat, broadened bands around \SI{1}{\tera\hertz} do not contribute significantly to transport because of their small group velocity and large linewidth.  This was confirmed by the temperature dependence of the spin Seebeck effect in GdIG/Pt bilayers \cite{SSEGdIG}. As detailed in Ref.~\cite{SSEGdIG}, the SSE can be understood considering that the transport of thermal magnons in GdIG is dominated by the  $\alpha$ and $\beta$ magnon modes.
At low temperatures only the $\alpha$-mode is thermally populated (red mode in Fig.~\ref{magnon_modes} (a)), while with increasing temperature the $\beta$-mode shifts below $k_\mathrm{B} T$ (blue mode in Fig.~\ref{magnon_modes} (c)) 
and dominates the SSE, leading to a low temperature SSE sign change \cite{SSEGdIG}. At $T_\mathrm{comp}$ the orientation of the magnetic sublattices is inverted and consequently the two magnon modes exchange roles (and polarization), as shown in Fig.~\ref{magnon_modes} (c) and (e), leading to an abrupt second sign change of the SSE signal at the compensation point \cite{SSEGdIG}. These sign changes can be understood considering that magnons with opposite polarization carry opposite angular momentum. The FMR-like magnons with positive polarization (red modes) reduce the net magnetization. The gapped magnons at higher frequencies arise from the precession of magnetic sublattice moments in the exchange field provided by the other sublattices, similar to the excitations in an antiferromagnet \cite{KefferKittelAFMR, Kittel_FerriMR}. Magnons with negative polarization (blue modes) correspond to the precession of sublattice moments aligned antiparallel to the external field, such that their excitation increases the net magnetization. The MMR arises from transfer of angular momentum from the electron to the magnon system (and vice versa) \cite{ZhangPRB}, implying that magnons with opposite polarization (carrying opposite angular momentum) may also lead to an opposite sign of the detected MMR signal. Since we observe no such sign change in the MMR in Fig.~\ref{SMR_MMR} (e), we conclude that either both modes contribute with the same sign or that the red mode dominates the MMR for the Pt strip separation of \SI{200}{\nano\meter} and the temperatures studied here.

\begin{figure}[!ht]
\includegraphics[width=\columnwidth]{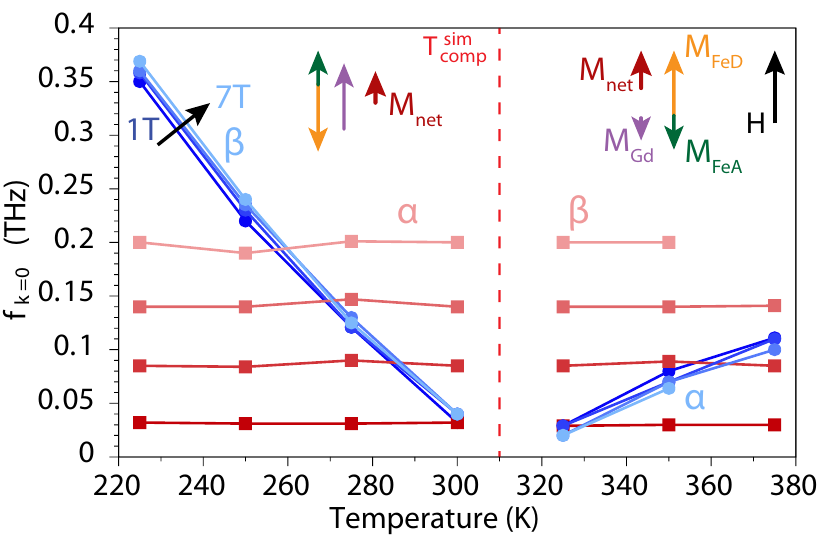}
\caption{Temperature and field dependence of the $k=0$ frequency for the $\alpha$ and $\beta$ modes. The red and blue color marks positive and negative polarization, respectively. The orientation of the sublattice and net magnetization direction with respect to the external field $\textbf{H}$ is depicted by arrows for temperatures above and below the compensation point $T_\mathrm{comp}^\mathrm{sim}=\SI{310}{\kelvin}$.}%
\label{fig:mode_dependence}%
\end{figure}

The contributions from different modes can be distinguished by studying the field and temperature dependence of the MMR. The frequency values at $k=0$ for the $\alpha$ and $\beta$ modes from Fig.~\ref{magnon_modes} are compiled in Fig.~\ref{fig:mode_dependence}. The orientation of the sublattice magnetizations with respect to the external field $\textbf{H}$ above and below the compensation point is indicated by arrows. 
Similar to the FMR-like mode in YIG, the fundamental (red) mode in Fig.~\ref{magnon_modes} shifts up in frequency proportional to the applied field as the Zeeman gap opens, but has a very weak temperature dependence (red symbols in Fig.~\ref{fig:mode_dependence}). The blue `exchange' mode has a much stronger temperature dependence (blue symbols in Fig.~\ref{fig:mode_dependence}) caused by relative changes in sublattice magnetization altering the effective exchange fields which determine this mode \cite{SSEGdIG}.

At low temperatures (Fig.~\ref{magnon_modes} (a) and (b)), the $\beta$-mode is not populated and does not contribute to the spin transport. The applied magnetic field freezes out the $\alpha$-magnons, thereby reducing the MMR signal (see Fig.~\ref{SMR_MMR} (e) and \ref{relsup}). With increasing temperature, the exchange gap of the $\beta$-mode decreases (see Fig.~\ref{magnon_modes} (c) and \ref{fig:mode_dependence}) due to the thermally induced disorder in the Gd system, leading to a shift of the exchange mode to lower frequencies \cite{SSEGdIG}. 
Below $T_\mathrm{comp}$, this trend is partly reverted by a magnetic field that increases the Gd order and thereby increases the $\beta$-mode frequencies again (see field dependence of the blue lines in Fig.~\ref{fig:mode_dependence}). The spin current carried by magnons with opposite polarization reduces the detected MMR amplitude. A magnetic field induced depopulation of the blue $\beta$ magnons below $T_\mathrm{comp}$ should therefore enhance the MMR signal. Although the frequency shift of the beta-mode predicted by simulations is small compared to the Zeeman shift of the alpha-mode, such a trend is indeed observed in the experimental MMR data displayed in Fig.~\ref{SMR_MMR} between $T_\mathrm{cross}$ and $T_\mathrm{comp}$.

Above the compensation temperature, the Gd moments are oriented antiparallel to the external field and depolarize when the latter is increased \cite{GanzhornPRB}, which implies a frequency reduction by magnetic field. In other words, the red (blue) mode shifts to higher (lower) frequencies and is depopulated (populated) with magnetic field. The observed strong suppression of the MMR above $T_\mathrm{comp}$ in Fig.~\ref{SMR_MMR} and \ref{relsup} is consistent with this picture. 

The study of the magnon spectrum therefore helps to understand the field and temperature dependence of the MMR in GdIG. While for a quantitative analysis additional factors such as the magnon transport properties need to be carefully considered, our results suggest that the magnon population plays an important role for the MMR effect.

In summary, we measured the non-local magnon mediated magnetoresistance effect (MMR) in the collinear as well as in the canted phase of the compensated ferrimagnet GdIG. The data taken close to the compensation temperature suggest that the MMR is suppressed in the canted phase, either due to inefficient magnon injection or to magnetic domain formation. The MMR signal is furthermore suppressed by magnetic field at low temperatures and above $T_\mathrm{comp}$, but is enhanced just below compensation. We relate these changes to the magnon spectrum of GdIG by comparison with atomistic simulations. We can qualitatively explain the field and temperature dependence of the MMR in terms of a competition between magnon modes with opposite polarization, confirming the importance of the magnon modes and their thermal population for spin transport in magnetic insulators.

This work is financially supported by the Deutsche Forschungsgemeinschaft through the Priority Program Spin Caloric Transport (GO 944/4, BA 2954/2 and GR 1132/18), JSPS KAKENHI Grant Nos. 25247056, 25220910, 26103006, a Laura-Bassi stipend of the Technische Universit\"at M\"unchen (to NV) and the Tohoku University Graduate Program in Spintronics.


\begin{thebibliography}{36}%
\makeatletter
\providecommand \@ifxundefined [1]{%
 \@ifx{#1\undefined}
}%
\providecommand \@ifnum [1]{%
 \ifnum #1\expandafter \@firstoftwo
 \else \expandafter \@secondoftwo
 \fi
}%
\providecommand \@ifx [1]{%
 \ifx #1\expandafter \@firstoftwo
 \else \expandafter \@secondoftwo
 \fi
}%
\providecommand \natexlab [1]{#1}%
\providecommand \enquote  [1]{``#1''}%
\providecommand \bibnamefont  [1]{#1}%
\providecommand \bibfnamefont [1]{#1}%
\providecommand \citenamefont [1]{#1}%
\providecommand \href@noop [0]{\@secondoftwo}%
\providecommand \href [0]{\begingroup \@sanitize@url \@href}%
\providecommand \@href[1]{\@@startlink{#1}\@@href}%
\providecommand \@@href[1]{\endgroup#1\@@endlink}%
\providecommand \@sanitize@url [0]{\catcode `\\12\catcode `\$12\catcode
  `\&12\catcode `\#12\catcode `\^12\catcode `\_12\catcode `\%12\relax}%
\providecommand \@@startlink[1]{}%
\providecommand \@@endlink[0]{}%
\providecommand \url  [0]{\begingroup\@sanitize@url \@url }%
\providecommand \@url [1]{\endgroup\@href {#1}{\urlprefix }}%
\providecommand \urlprefix  [0]{URL }%
\providecommand \Eprint [0]{\href }%
\providecommand \doibase [0]{http://dx.doi.org/}%
\providecommand \selectlanguage [0]{\@gobble}%
\providecommand \bibinfo  [0]{\@secondoftwo}%
\providecommand \bibfield  [0]{\@secondoftwo}%
\providecommand \translation [1]{[#1]}%
\providecommand \BibitemOpen [0]{}%
\providecommand \bibitemStop [0]{}%
\providecommand \bibitemNoStop [0]{.\EOS\space}%
\providecommand \EOS [0]{\spacefactor3000\relax}%
\providecommand \BibitemShut  [1]{\csname bibitem#1\endcsname}%
\let\auto@bib@innerbib\@empty
\bibitem [{\citenamefont {Cornelissen}\ \emph {et~al.}(2015)\citenamefont
  {Cornelissen}, \citenamefont {Liu}, \citenamefont {Duine}, \citenamefont
  {Ben~Youssef},\ and\ \citenamefont {van Wees}}]{CornelissenNat2015}%
  \BibitemOpen
  \bibfield  {author} {\bibinfo {author} {\bibfnamefont {L.~J.}\ \bibnamefont
  {Cornelissen}}, \bibinfo {author} {\bibfnamefont {J.}~\bibnamefont {Liu}},
  \bibinfo {author} {\bibfnamefont {R.~A.}\ \bibnamefont {Duine}}, \bibinfo
  {author} {\bibfnamefont {J.}~\bibnamefont {Ben~Youssef}}, \ and\ \bibinfo
  {author} {\bibfnamefont {B.~J.}\ \bibnamefont {van Wees}},\ }\href@noop {}
  {\bibfield  {journal} {\bibinfo  {journal} {Nat. Phys.}\ }\textbf {\bibinfo
  {volume} {11}},\ \bibinfo {pages} {1022} (\bibinfo {year}
  {2015})}\BibitemShut {NoStop}%
\bibitem [{\citenamefont {Goennenwein}\ \emph {et~al.}(2015)\citenamefont
  {Goennenwein}, \citenamefont {Schlitz}, \citenamefont {Pernpeintner},
  \citenamefont {Ganzhorn}, \citenamefont {Althammer}, \citenamefont {Gross},\
  and\ \citenamefont {Huebl}}]{Goennenwein2015}%
  \BibitemOpen
  \bibfield  {author} {\bibinfo {author} {\bibfnamefont {S.~T.~B.}\
  \bibnamefont {Goennenwein}}, \bibinfo {author} {\bibfnamefont
  {R.}~\bibnamefont {Schlitz}}, \bibinfo {author} {\bibfnamefont
  {M.}~\bibnamefont {Pernpeintner}}, \bibinfo {author} {\bibfnamefont
  {K.}~\bibnamefont {Ganzhorn}}, \bibinfo {author} {\bibfnamefont
  {M.}~\bibnamefont {Althammer}}, \bibinfo {author} {\bibfnamefont
  {R.}~\bibnamefont {Gross}}, \ and\ \bibinfo {author} {\bibfnamefont
  {H.}~\bibnamefont {Huebl}},\ }\href {\doibase 10.1063/1.4935074} {\bibfield
  {journal} {\bibinfo  {journal} {Appl. Phys. Lett.}\ }\textbf {\bibinfo
  {volume} {107}},\ \bibinfo {pages} {172405} (\bibinfo {year}
  {2015})}\BibitemShut {NoStop}%
\bibitem [{\citenamefont {Zhang}\ and\ \citenamefont
  {Zhang}(2012{\natexlab{a}})}]{ZhangPRL}%
  \BibitemOpen
  \bibfield  {author} {\bibinfo {author} {\bibfnamefont {S.~S.-L.}\
  \bibnamefont {Zhang}}\ and\ \bibinfo {author} {\bibfnamefont
  {S.}~\bibnamefont {Zhang}},\ }\href {\doibase 10.1103/PhysRevLett.109.096603}
  {\bibfield  {journal} {\bibinfo  {journal} {Phys. Rev. Lett.}\ }\textbf
  {\bibinfo {volume} {109}},\ \bibinfo {pages} {096603} (\bibinfo {year}
  {2012}{\natexlab{a}})}\BibitemShut {NoStop}%
\bibitem [{\citenamefont {Cornelissen}\ \emph
  {et~al.}(2016{\natexlab{a}})\citenamefont {Cornelissen}, \citenamefont
  {Peters}, \citenamefont {Bauer}, \citenamefont {Duine},\ and\ \citenamefont
  {van Wees}}]{CornelissenPRB94}%
  \BibitemOpen
  \bibfield  {author} {\bibinfo {author} {\bibfnamefont {L.~J.}\ \bibnamefont
  {Cornelissen}}, \bibinfo {author} {\bibfnamefont {K.~J.~H.}\ \bibnamefont
  {Peters}}, \bibinfo {author} {\bibfnamefont {G.~E.~W.}\ \bibnamefont
  {Bauer}}, \bibinfo {author} {\bibfnamefont {R.~A.}\ \bibnamefont {Duine}}, \
  and\ \bibinfo {author} {\bibfnamefont {B.~J.}\ \bibnamefont {van Wees}},\
  }\href {\doibase 10.1103/PhysRevB.94.014412} {\bibfield  {journal} {\bibinfo
  {journal} {Phys. Rev. B}\ }\textbf {\bibinfo {volume} {94}},\ \bibinfo
  {pages} {014412} (\bibinfo {year} {2016}{\natexlab{a}})}\BibitemShut
  {NoStop}%
\bibitem [{\citenamefont {V\'elez}\ \emph {et~al.}(2016)\citenamefont
  {V\'elez}, \citenamefont {Bedoya-Pinto}, \citenamefont {Yan}, \citenamefont
  {Hueso},\ and\ \citenamefont {Casanova}}]{VelezMMR}%
  \BibitemOpen
  \bibfield  {author} {\bibinfo {author} {\bibfnamefont {S.}~\bibnamefont
  {V\'elez}}, \bibinfo {author} {\bibfnamefont {A.}~\bibnamefont
  {Bedoya-Pinto}}, \bibinfo {author} {\bibfnamefont {W.}~\bibnamefont {Yan}},
  \bibinfo {author} {\bibfnamefont {L.~E.}\ \bibnamefont {Hueso}}, \ and\
  \bibinfo {author} {\bibfnamefont {F.}~\bibnamefont {Casanova}},\ }\href
  {\doibase 10.1103/PhysRevB.94.174405} {\bibfield  {journal} {\bibinfo
  {journal} {Phys. Rev. B}\ }\textbf {\bibinfo {volume} {94}},\ \bibinfo
  {pages} {174405} (\bibinfo {year} {2016})}\BibitemShut {NoStop}%
\bibitem [{\citenamefont {Li}\ \emph {et~al.}(2016)\citenamefont {Li},
  \citenamefont {Xu}, \citenamefont {Aldosary}, \citenamefont {Tang},
  \citenamefont {Lin}, \citenamefont {Zhang}, \citenamefont {Lake},\ and\
  \citenamefont {Shi}}]{Li_MMR_trilayer}%
  \BibitemOpen
  \bibfield  {author} {\bibinfo {author} {\bibfnamefont {J.}~\bibnamefont
  {Li}}, \bibinfo {author} {\bibfnamefont {Y.}~\bibnamefont {Xu}}, \bibinfo
  {author} {\bibfnamefont {M.}~\bibnamefont {Aldosary}}, \bibinfo {author}
  {\bibfnamefont {C.}~\bibnamefont {Tang}}, \bibinfo {author} {\bibfnamefont
  {Z.}~\bibnamefont {Lin}}, \bibinfo {author} {\bibfnamefont {S.}~\bibnamefont
  {Zhang}}, \bibinfo {author} {\bibfnamefont {R.}~\bibnamefont {Lake}}, \ and\
  \bibinfo {author} {\bibfnamefont {J.}~\bibnamefont {Shi}},\ }\href
  {http://dx.doi.org/10.1038/ncomms10858} {\bibfield  {journal} {\bibinfo
  {journal} {Nat. Commun.}\ }\textbf {\bibinfo {volume} {7}},\ \bibinfo {pages}
  {10858} (\bibinfo {year} {2016})}\BibitemShut {NoStop}%
\bibitem [{\citenamefont {Wu}\ \emph {et~al.}(2016)\citenamefont {Wu},
  \citenamefont {Wan}, \citenamefont {Zhang}, \citenamefont {Yuan},
  \citenamefont {Zhang}, \citenamefont {Qin}, \citenamefont {Wei},
  \citenamefont {Han},\ and\ \citenamefont {Zhang}}]{Wu_MMR}%
  \BibitemOpen
  \bibfield  {author} {\bibinfo {author} {\bibfnamefont {H.}~\bibnamefont
  {Wu}}, \bibinfo {author} {\bibfnamefont {C.~H.}\ \bibnamefont {Wan}},
  \bibinfo {author} {\bibfnamefont {X.}~\bibnamefont {Zhang}}, \bibinfo
  {author} {\bibfnamefont {Z.~H.}\ \bibnamefont {Yuan}}, \bibinfo {author}
  {\bibfnamefont {Q.~T.}\ \bibnamefont {Zhang}}, \bibinfo {author}
  {\bibfnamefont {J.~Y.}\ \bibnamefont {Qin}}, \bibinfo {author} {\bibfnamefont
  {H.~X.}\ \bibnamefont {Wei}}, \bibinfo {author} {\bibfnamefont {X.~F.}\
  \bibnamefont {Han}}, \ and\ \bibinfo {author} {\bibfnamefont
  {S.}~\bibnamefont {Zhang}},\ }\href {\doibase 10.1103/PhysRevB.93.060403}
  {\bibfield  {journal} {\bibinfo  {journal} {Phys. Rev. B}\ }\textbf {\bibinfo
  {volume} {93}},\ \bibinfo {pages} {060403} (\bibinfo {year}
  {2016})}\BibitemShut {NoStop}%
\bibitem [{\citenamefont {Dyakonov}\ and\ \citenamefont
  {Perel}(1971)}]{Dyakonov}%
  \BibitemOpen
  \bibfield  {author} {\bibinfo {author} {\bibfnamefont {M.}~\bibnamefont
  {Dyakonov}}\ and\ \bibinfo {author} {\bibfnamefont {V.}~\bibnamefont
  {Perel}},\ }\href {http://www.jetpletters.ac.ru/ps/1587/index.shtml}
  {\bibfield  {journal} {\bibinfo  {journal} {JETP Lett.}\ }\textbf {\bibinfo
  {volume} {13}},\ \bibinfo {pages} {467} (\bibinfo {year} {1971})}\BibitemShut
  {NoStop}%
\bibitem [{\citenamefont {Hirsch}(1999)}]{Hirsch}%
  \BibitemOpen
  \bibfield  {author} {\bibinfo {author} {\bibfnamefont {J.~E.}\ \bibnamefont
  {Hirsch}},\ }\href {\doibase 10.1103/PhysRevLett.83.1834} {\bibfield
  {journal} {\bibinfo  {journal} {Phys. Rev. Lett.}\ }\textbf {\bibinfo
  {volume} {83}},\ \bibinfo {pages} {1834} (\bibinfo {year}
  {1999})}\BibitemShut {NoStop}%
\bibitem [{\citenamefont {Cornelissen}\ and\ \citenamefont {van
  Wees}(2016)}]{CornelissenField}%
  \BibitemOpen
  \bibfield  {author} {\bibinfo {author} {\bibfnamefont {L.~J.}\ \bibnamefont
  {Cornelissen}}\ and\ \bibinfo {author} {\bibfnamefont {B.~J.}\ \bibnamefont
  {van Wees}},\ }\href {\doibase 10.1103/PhysRevB.93.020403} {\bibfield
  {journal} {\bibinfo  {journal} {Phys. Rev. B}\ }\textbf {\bibinfo {volume}
  {93}},\ \bibinfo {pages} {020403} (\bibinfo {year} {2016})}\BibitemShut
  {NoStop}%
\bibitem [{\citenamefont {Cornelissen}\ \emph
  {et~al.}(2016{\natexlab{b}})\citenamefont {Cornelissen}, \citenamefont
  {Shan},\ and\ \citenamefont {van Wees}}]{CornelissenPRB94.180402}%
  \BibitemOpen
  \bibfield  {author} {\bibinfo {author} {\bibfnamefont {L.~J.}\ \bibnamefont
  {Cornelissen}}, \bibinfo {author} {\bibfnamefont {J.}~\bibnamefont {Shan}}, \
  and\ \bibinfo {author} {\bibfnamefont {B.~J.}\ \bibnamefont {van Wees}},\
  }\href {\doibase 10.1103/PhysRevB.94.180402} {\bibfield  {journal} {\bibinfo
  {journal} {Phys. Rev. B}\ }\textbf {\bibinfo {volume} {94}},\ \bibinfo
  {pages} {180402} (\bibinfo {year} {2016}{\natexlab{b}})}\BibitemShut
  {NoStop}%
\bibitem [{\citenamefont {Shan}\ \emph {et~al.}(2017)\citenamefont {Shan},
  \citenamefont {Bougiatioti}, \citenamefont {Liang}, \citenamefont {Reiss},
  \citenamefont {Kuschel},\ and\ \citenamefont {van Wees}}]{ShanNFO}%
  \BibitemOpen
  \bibfield  {author} {\bibinfo {author} {\bibfnamefont {J.}~\bibnamefont
  {Shan}}, \bibinfo {author} {\bibfnamefont {P.}~\bibnamefont {Bougiatioti}},
  \bibinfo {author} {\bibfnamefont {L.}~\bibnamefont {Liang}}, \bibinfo
  {author} {\bibfnamefont {G.}~\bibnamefont {Reiss}}, \bibinfo {author}
  {\bibfnamefont {T.}~\bibnamefont {Kuschel}}, \ and\ \bibinfo {author}
  {\bibfnamefont {B.~J.}\ \bibnamefont {van Wees}},\ }\href {\doibase
  10.1063/1.4979408} {\bibfield  {journal} {\bibinfo  {journal} {Appl. Phys.
  Lett.}\ }\textbf {\bibinfo {volume} {110}},\ \bibinfo {pages} {132406}
  (\bibinfo {year} {2017})},\ \Eprint
  {http://arxiv.org/abs/http://dx.doi.org/10.1063/1.4979408}
  {http://dx.doi.org/10.1063/1.4979408} \BibitemShut {NoStop}%
\bibitem [{\citenamefont {Dionne}(2009)}]{Dionne:Magnetic:Oxides:Book:2009}%
  \BibitemOpen
  \bibfield  {author} {\bibinfo {author} {\bibfnamefont {G.~F.}\ \bibnamefont
  {Dionne}},\ }\href@noop {} {\emph {\bibinfo {title} {Magnetic Oxides}}}\
  (\bibinfo  {publisher} {Springer},\ \bibinfo {address} {New York},\ \bibinfo
  {year} {2009})\BibitemShut {NoStop}%
\bibitem [{\citenamefont {Ganzhorn}\ \emph {et~al.}(2016)\citenamefont
  {Ganzhorn}, \citenamefont {Barker}, \citenamefont {Schlitz}, \citenamefont
  {Piot}, \citenamefont {Ollefs}, \citenamefont {Guillou}, \citenamefont
  {Wilhelm}, \citenamefont {Rogalev}, \citenamefont {Opel}, \citenamefont
  {Althammer}, \citenamefont {Gepr\"ags}, \citenamefont {Huebl}, \citenamefont
  {Gross}, \citenamefont {Bauer},\ and\ \citenamefont
  {Goennenwein}}]{GanzhornPRB}%
  \BibitemOpen
  \bibfield  {author} {\bibinfo {author} {\bibfnamefont {K.}~\bibnamefont
  {Ganzhorn}}, \bibinfo {author} {\bibfnamefont {J.}~\bibnamefont {Barker}},
  \bibinfo {author} {\bibfnamefont {R.}~\bibnamefont {Schlitz}}, \bibinfo
  {author} {\bibfnamefont {B.~A.}\ \bibnamefont {Piot}}, \bibinfo {author}
  {\bibfnamefont {K.}~\bibnamefont {Ollefs}}, \bibinfo {author} {\bibfnamefont
  {F.}~\bibnamefont {Guillou}}, \bibinfo {author} {\bibfnamefont
  {F.}~\bibnamefont {Wilhelm}}, \bibinfo {author} {\bibfnamefont
  {A.}~\bibnamefont {Rogalev}}, \bibinfo {author} {\bibfnamefont
  {M.}~\bibnamefont {Opel}}, \bibinfo {author} {\bibfnamefont {M.}~\bibnamefont
  {Althammer}}, \bibinfo {author} {\bibfnamefont {S.}~\bibnamefont
  {Gepr\"ags}}, \bibinfo {author} {\bibfnamefont {H.}~\bibnamefont {Huebl}},
  \bibinfo {author} {\bibfnamefont {R.}~\bibnamefont {Gross}}, \bibinfo
  {author} {\bibfnamefont {G.~E.~W.}\ \bibnamefont {Bauer}}, \ and\ \bibinfo
  {author} {\bibfnamefont {S.~T.~B.}\ \bibnamefont {Goennenwein}},\ }\href
  {\doibase 10.1103/PhysRevB.94.094401} {\bibfield  {journal} {\bibinfo
  {journal} {Phys. Rev. B}\ }\textbf {\bibinfo {volume} {94}},\ \bibinfo
  {pages} {094401} (\bibinfo {year} {2016})}\BibitemShut {NoStop}%
\bibitem [{\citenamefont {Bernasconi}\ and\ \citenamefont
  {Kuse}(1971)}]{Bernasconi}%
  \BibitemOpen
  \bibfield  {author} {\bibinfo {author} {\bibfnamefont {J.}~\bibnamefont
  {Bernasconi}}\ and\ \bibinfo {author} {\bibfnamefont {D.}~\bibnamefont
  {Kuse}},\ }\href {\doibase 10.1103/PhysRevB.3.811} {\bibfield  {journal}
  {\bibinfo  {journal} {Phys. Rev. B}\ }\textbf {\bibinfo {volume} {3}},\
  \bibinfo {pages} {811} (\bibinfo {year} {1971})}\BibitemShut {NoStop}%
\bibitem [{\citenamefont {Clark}\ and\ \citenamefont {Callen}(1968)}]{Clark}%
  \BibitemOpen
  \bibfield  {author} {\bibinfo {author} {\bibfnamefont {A.~E.}\ \bibnamefont
  {Clark}}\ and\ \bibinfo {author} {\bibfnamefont {E.}~\bibnamefont {Callen}},\
  }\href {\doibase 10.1063/1.1656100} {\bibfield  {journal} {\bibinfo
  {journal} {J. Appl. Phys.}\ }\textbf {\bibinfo {volume} {39}},\ \bibinfo
  {pages} {5972} (\bibinfo {year} {1968})}\BibitemShut {NoStop}%
\bibitem [{\citenamefont {Althammer}\ \emph {et~al.}(2013)\citenamefont
  {Althammer}, \citenamefont {Meyer}, \citenamefont {Nakayama}, \citenamefont
  {Schreier}, \citenamefont {Altmannshofer}, \citenamefont {Weiler},
  \citenamefont {Huebl}, \citenamefont {Gepr\"ags}, \citenamefont {Opel},
  \citenamefont {Gross}, \citenamefont {Meier}, \citenamefont {Klewe},
  \citenamefont {Kuschel}, \citenamefont {Schmalhorst}, \citenamefont {Reiss},
  \citenamefont {Shen}, \citenamefont {Gupta}, \citenamefont {Chen},
  \citenamefont {Bauer}, \citenamefont {Saitoh},\ and\ \citenamefont
  {Goennenwein}}]{AltiPRB}%
  \BibitemOpen
  \bibfield  {author} {\bibinfo {author} {\bibfnamefont {M.}~\bibnamefont
  {Althammer}}, \bibinfo {author} {\bibfnamefont {S.}~\bibnamefont {Meyer}},
  \bibinfo {author} {\bibfnamefont {H.}~\bibnamefont {Nakayama}}, \bibinfo
  {author} {\bibfnamefont {M.}~\bibnamefont {Schreier}}, \bibinfo {author}
  {\bibfnamefont {S.}~\bibnamefont {Altmannshofer}}, \bibinfo {author}
  {\bibfnamefont {M.}~\bibnamefont {Weiler}}, \bibinfo {author} {\bibfnamefont
  {H.}~\bibnamefont {Huebl}}, \bibinfo {author} {\bibfnamefont
  {S.}~\bibnamefont {Gepr\"ags}}, \bibinfo {author} {\bibfnamefont
  {M.}~\bibnamefont {Opel}}, \bibinfo {author} {\bibfnamefont {R.}~\bibnamefont
  {Gross}}, \bibinfo {author} {\bibfnamefont {D.}~\bibnamefont {Meier}},
  \bibinfo {author} {\bibfnamefont {C.}~\bibnamefont {Klewe}}, \bibinfo
  {author} {\bibfnamefont {T.}~\bibnamefont {Kuschel}}, \bibinfo {author}
  {\bibfnamefont {J.-M.}\ \bibnamefont {Schmalhorst}}, \bibinfo {author}
  {\bibfnamefont {G.}~\bibnamefont {Reiss}}, \bibinfo {author} {\bibfnamefont
  {L.}~\bibnamefont {Shen}}, \bibinfo {author} {\bibfnamefont {A.}~\bibnamefont
  {Gupta}}, \bibinfo {author} {\bibfnamefont {Y.-T.}\ \bibnamefont {Chen}},
  \bibinfo {author} {\bibfnamefont {G.~E.~W.}\ \bibnamefont {Bauer}}, \bibinfo
  {author} {\bibfnamefont {E.}~\bibnamefont {Saitoh}}, \ and\ \bibinfo {author}
  {\bibfnamefont {S.~T.~B.}\ \bibnamefont {Goennenwein}},\ }\href {\doibase
  10.1103/PhysRevB.87.224401} {\bibfield  {journal} {\bibinfo  {journal} {Phys.
  Rev. B}\ }\textbf {\bibinfo {volume} {87}},\ \bibinfo {pages} {224401}
  (\bibinfo {year} {2013})}\BibitemShut {NoStop}%
\bibitem [{\citenamefont {Chen}\ \emph {et~al.}(2013)\citenamefont {Chen},
  \citenamefont {Takahashi}, \citenamefont {Nakayama}, \citenamefont
  {Althammer}, \citenamefont {Goennenwein}, \citenamefont {Saitoh},\ and\
  \citenamefont {Bauer}}]{ChenSMR}%
  \BibitemOpen
  \bibfield  {author} {\bibinfo {author} {\bibfnamefont {Y.-T.}\ \bibnamefont
  {Chen}}, \bibinfo {author} {\bibfnamefont {S.}~\bibnamefont {Takahashi}},
  \bibinfo {author} {\bibfnamefont {H.}~\bibnamefont {Nakayama}}, \bibinfo
  {author} {\bibfnamefont {M.}~\bibnamefont {Althammer}}, \bibinfo {author}
  {\bibfnamefont {S.~T.~B.}\ \bibnamefont {Goennenwein}}, \bibinfo {author}
  {\bibfnamefont {E.}~\bibnamefont {Saitoh}}, \ and\ \bibinfo {author}
  {\bibfnamefont {G.~E.~W.}\ \bibnamefont {Bauer}},\ }\href {\doibase
  10.1103/PhysRevB.87.144411} {\bibfield  {journal} {\bibinfo  {journal} {Phys.
  Rev. B}\ }\textbf {\bibinfo {volume} {87}},\ \bibinfo {pages} {144411}
  (\bibinfo {year} {2013})}\BibitemShut {NoStop}%
\bibitem [{\citenamefont {Nakayama}\ \emph {et~al.}(2013)\citenamefont
  {Nakayama}, \citenamefont {Althammer}, \citenamefont {Chen}, \citenamefont
  {Uchida}, \citenamefont {Kajiwara}, \citenamefont {Kikuchi}, \citenamefont
  {Ohtani}, \citenamefont {Gepr\"ags}, \citenamefont {Opel}, \citenamefont
  {Takahashi}, \citenamefont {Gross}, \citenamefont {Bauer}, \citenamefont
  {Goennenwein},\ and\ \citenamefont {Saitoh}}]{NakayamaSMR}%
  \BibitemOpen
  \bibfield  {author} {\bibinfo {author} {\bibfnamefont {H.}~\bibnamefont
  {Nakayama}}, \bibinfo {author} {\bibfnamefont {M.}~\bibnamefont {Althammer}},
  \bibinfo {author} {\bibfnamefont {Y.-T.}\ \bibnamefont {Chen}}, \bibinfo
  {author} {\bibfnamefont {K.}~\bibnamefont {Uchida}}, \bibinfo {author}
  {\bibfnamefont {Y.}~\bibnamefont {Kajiwara}}, \bibinfo {author}
  {\bibfnamefont {D.}~\bibnamefont {Kikuchi}}, \bibinfo {author} {\bibfnamefont
  {T.}~\bibnamefont {Ohtani}}, \bibinfo {author} {\bibfnamefont
  {S.}~\bibnamefont {Gepr\"ags}}, \bibinfo {author} {\bibfnamefont
  {M.}~\bibnamefont {Opel}}, \bibinfo {author} {\bibfnamefont {S.}~\bibnamefont
  {Takahashi}}, \bibinfo {author} {\bibfnamefont {R.}~\bibnamefont {Gross}},
  \bibinfo {author} {\bibfnamefont {G.~E.~W.}\ \bibnamefont {Bauer}}, \bibinfo
  {author} {\bibfnamefont {S.~T.~B.}\ \bibnamefont {Goennenwein}}, \ and\
  \bibinfo {author} {\bibfnamefont {E.}~\bibnamefont {Saitoh}},\ }\href
  {\doibase 10.1103/PhysRevLett.110.206601} {\bibfield  {journal} {\bibinfo
  {journal} {Phys. Rev. Lett.}\ }\textbf {\bibinfo {volume} {110}},\ \bibinfo
  {pages} {206601} (\bibinfo {year} {2013})}\BibitemShut {NoStop}%
\bibitem [{\citenamefont {Vlietstra}\ \emph {et~al.}(2013)\citenamefont
  {Vlietstra}, \citenamefont {Shan}, \citenamefont {Castel}, \citenamefont {van
  Wees},\ and\ \citenamefont {Ben~Youssef}}]{SMR:Vlietstra:PRB:2013}%
  \BibitemOpen
  \bibfield  {author} {\bibinfo {author} {\bibfnamefont {N.}~\bibnamefont
  {Vlietstra}}, \bibinfo {author} {\bibfnamefont {J.}~\bibnamefont {Shan}},
  \bibinfo {author} {\bibfnamefont {V.}~\bibnamefont {Castel}}, \bibinfo
  {author} {\bibfnamefont {B.~J.}\ \bibnamefont {van Wees}}, \ and\ \bibinfo
  {author} {\bibfnamefont {J.}~\bibnamefont {Ben~Youssef}},\ }\href {\doibase
  10.1103/PhysRevB.87.184421} {\bibfield  {journal} {\bibinfo  {journal} {Phys.
  Rev. B}\ }\textbf {\bibinfo {volume} {87}},\ \bibinfo {pages} {184421}
  (\bibinfo {year} {2013})}\BibitemShut {NoStop}%
\bibitem [{\citenamefont {Aqeel}\ \emph {et~al.}(2016)\citenamefont {Aqeel},
  \citenamefont {Vlietstra}, \citenamefont {Roy}, \citenamefont {Mostovoy},
  \citenamefont {van Wees},\ and\ \citenamefont {Palstra}}]{AqeelCSO}%
  \BibitemOpen
  \bibfield  {author} {\bibinfo {author} {\bibfnamefont {A.}~\bibnamefont
  {Aqeel}}, \bibinfo {author} {\bibfnamefont {N.}~\bibnamefont {Vlietstra}},
  \bibinfo {author} {\bibfnamefont {A.}~\bibnamefont {Roy}}, \bibinfo {author}
  {\bibfnamefont {M.}~\bibnamefont {Mostovoy}}, \bibinfo {author}
  {\bibfnamefont {B.~J.}\ \bibnamefont {van Wees}}, \ and\ \bibinfo {author}
  {\bibfnamefont {T.~T.~M.}\ \bibnamefont {Palstra}},\ }\href {\doibase
  10.1103/PhysRevB.94.134418} {\bibfield  {journal} {\bibinfo  {journal} {Phys.
  Rev. B}\ }\textbf {\bibinfo {volume} {94}},\ \bibinfo {pages} {134418}
  (\bibinfo {year} {2016})}\BibitemShut {NoStop}%
\bibitem [{\citenamefont {Gepr\"{a}gs}\ \emph {et~al.}(2016)\citenamefont
  {Gepr\"{a}gs}, \citenamefont {Kehlberger}, \citenamefont {Coletta},
  \citenamefont {Qiu}, \citenamefont {Guo}, \citenamefont {Schulz},
  \citenamefont {Mix}, \citenamefont {Meyer}, \citenamefont {Kamra},
  \citenamefont {Althammer}, \citenamefont {Huebl}, \citenamefont {Jakob},
  \citenamefont {Ohnuma}, \citenamefont {Adachi}, \citenamefont {Barker},
  \citenamefont {Maekawa}, \citenamefont {Bauer}, \citenamefont {Saitoh},
  \citenamefont {Gross}, \citenamefont {Goennenwein},\ and\ \citenamefont
  {Kl\"{a}ui}}]{SSEGdIG}%
  \BibitemOpen
  \bibfield  {author} {\bibinfo {author} {\bibfnamefont {S.}~\bibnamefont
  {Gepr\"{a}gs}}, \bibinfo {author} {\bibfnamefont {A.}~\bibnamefont
  {Kehlberger}}, \bibinfo {author} {\bibfnamefont {F.~D.}\ \bibnamefont
  {Coletta}}, \bibinfo {author} {\bibfnamefont {Z.}~\bibnamefont {Qiu}},
  \bibinfo {author} {\bibfnamefont {E.-J.}\ \bibnamefont {Guo}}, \bibinfo
  {author} {\bibfnamefont {T.}~\bibnamefont {Schulz}}, \bibinfo {author}
  {\bibfnamefont {C.}~\bibnamefont {Mix}}, \bibinfo {author} {\bibfnamefont
  {S.}~\bibnamefont {Meyer}}, \bibinfo {author} {\bibfnamefont
  {A.}~\bibnamefont {Kamra}}, \bibinfo {author} {\bibfnamefont
  {M.}~\bibnamefont {Althammer}}, \bibinfo {author} {\bibfnamefont
  {H.}~\bibnamefont {Huebl}}, \bibinfo {author} {\bibfnamefont
  {G.}~\bibnamefont {Jakob}}, \bibinfo {author} {\bibfnamefont
  {Y.}~\bibnamefont {Ohnuma}}, \bibinfo {author} {\bibfnamefont
  {H.}~\bibnamefont {Adachi}}, \bibinfo {author} {\bibfnamefont
  {J.}~\bibnamefont {Barker}}, \bibinfo {author} {\bibfnamefont
  {S.}~\bibnamefont {Maekawa}}, \bibinfo {author} {\bibfnamefont {G.~E.~W.}\
  \bibnamefont {Bauer}}, \bibinfo {author} {\bibfnamefont {E.}~\bibnamefont
  {Saitoh}}, \bibinfo {author} {\bibfnamefont {R.}~\bibnamefont {Gross}},
  \bibinfo {author} {\bibfnamefont {S.~T.~B.}\ \bibnamefont {Goennenwein}}, \
  and\ \bibinfo {author} {\bibfnamefont {M.}~\bibnamefont {Kl\"{a}ui}},\ }\href
  {\doibase 10.1038/ncomms10452} {\bibfield  {journal} {\bibinfo  {journal}
  {Nat. Commun.}\ }\textbf {\bibinfo {volume} {7}},\ \bibinfo {pages} {10452}
  (\bibinfo {year} {2016})}\BibitemShut {NoStop}%
\bibitem [{\citenamefont {Cramer}\ \emph {et~al.}(2017)\citenamefont {Cramer},
  \citenamefont {Guo}, \citenamefont {Gepr\"ags}, \citenamefont {Kehlberger},
  \citenamefont {Ivanov}, \citenamefont {Ganzhorn}, \citenamefont
  {Della~Coletta}, \citenamefont {Althammer}, \citenamefont {Huebl},
  \citenamefont {Gross}, \citenamefont {Kosel}, \citenamefont {Kl\"aui},\ and\
  \citenamefont {Goennenwein}}]{CramerSSE}%
  \BibitemOpen
  \bibfield  {author} {\bibinfo {author} {\bibfnamefont {J.}~\bibnamefont
  {Cramer}}, \bibinfo {author} {\bibfnamefont {E.-J.}\ \bibnamefont {Guo}},
  \bibinfo {author} {\bibfnamefont {S.}~\bibnamefont {Gepr\"ags}}, \bibinfo
  {author} {\bibfnamefont {A.}~\bibnamefont {Kehlberger}}, \bibinfo {author}
  {\bibfnamefont {Y.~P.}\ \bibnamefont {Ivanov}}, \bibinfo {author}
  {\bibfnamefont {K.}~\bibnamefont {Ganzhorn}}, \bibinfo {author}
  {\bibfnamefont {F.}~\bibnamefont {Della~Coletta}}, \bibinfo {author}
  {\bibfnamefont {M.}~\bibnamefont {Althammer}}, \bibinfo {author}
  {\bibfnamefont {H.}~\bibnamefont {Huebl}}, \bibinfo {author} {\bibfnamefont
  {R.}~\bibnamefont {Gross}}, \bibinfo {author} {\bibfnamefont
  {J.}~\bibnamefont {Kosel}}, \bibinfo {author} {\bibfnamefont
  {M.}~\bibnamefont {Kl\"aui}}, \ and\ \bibinfo {author} {\bibfnamefont
  {S.~T.~B.}\ \bibnamefont {Goennenwein}},\ }\href {\doibase
  10.1021/acs.nanolett.6b04522} {\bibfield  {journal} {\bibinfo  {journal}
  {Nano Letters}\ }\textbf {\bibinfo {volume} {0}},\ \bibinfo {pages} {null}
  (\bibinfo {year} {2017})},\ \bibinfo {note} {pMID: 28406308},\ \Eprint
  {http://arxiv.org/abs/http://dx.doi.org/10.1021/acs.nanolett.6b04522}
  {http://dx.doi.org/10.1021/acs.nanolett.6b04522} \BibitemShut {NoStop}%
\bibitem [{\citenamefont {P\"utter}\ \emph {et~al.}(2017)\citenamefont
  {P\"utter}, \citenamefont {Gepr\"ags}, \citenamefont {Schlitz}, \citenamefont
  {Althammer}, \citenamefont {Erb}, \citenamefont {Gross},\ and\ \citenamefont
  {Goennenwein}}]{PuetterSMR}%
  \BibitemOpen
  \bibfield  {author} {\bibinfo {author} {\bibfnamefont {S.}~\bibnamefont
  {P\"utter}}, \bibinfo {author} {\bibfnamefont {S.}~\bibnamefont {Gepr\"ags}},
  \bibinfo {author} {\bibfnamefont {R.}~\bibnamefont {Schlitz}}, \bibinfo
  {author} {\bibfnamefont {M.}~\bibnamefont {Althammer}}, \bibinfo {author}
  {\bibfnamefont {A.}~\bibnamefont {Erb}}, \bibinfo {author} {\bibfnamefont
  {R.}~\bibnamefont {Gross}}, \ and\ \bibinfo {author} {\bibfnamefont
  {S.~T.~B.}\ \bibnamefont {Goennenwein}},\ }\href {\doibase 10.1063/1.4973460}
  {\bibfield  {journal} {\bibinfo  {journal} {Appl. Phys. Lett.}\ }\textbf
  {\bibinfo {volume} {110}},\ \bibinfo {pages} {012403} (\bibinfo {year}
  {2017})},\ \Eprint {http://arxiv.org/abs/http://dx.doi.org/10.1063/1.4973460}
  {http://dx.doi.org/10.1063/1.4973460} \BibitemShut {NoStop}%
\bibitem [{\citenamefont {Schreier}\ \emph {et~al.}(2013)\citenamefont
  {Schreier}, \citenamefont {Roschewsky}, \citenamefont {Dobler}, \citenamefont
  {Meyer}, \citenamefont {Huebl}, \citenamefont {Gross},\ and\ \citenamefont
  {Goennenwein}}]{Schreier_iSSE}%
  \BibitemOpen
  \bibfield  {author} {\bibinfo {author} {\bibfnamefont {M.}~\bibnamefont
  {Schreier}}, \bibinfo {author} {\bibfnamefont {N.}~\bibnamefont
  {Roschewsky}}, \bibinfo {author} {\bibfnamefont {E.}~\bibnamefont {Dobler}},
  \bibinfo {author} {\bibfnamefont {S.}~\bibnamefont {Meyer}}, \bibinfo
  {author} {\bibfnamefont {H.}~\bibnamefont {Huebl}}, \bibinfo {author}
  {\bibfnamefont {R.}~\bibnamefont {Gross}}, \ and\ \bibinfo {author}
  {\bibfnamefont {S.~T.~B.}\ \bibnamefont {Goennenwein}},\ }\href {\doibase
  http://dx.doi.org/10.1063/1.4839395} {\bibfield  {journal} {\bibinfo
  {journal} {Appl. Phys. Lett.}\ }\textbf {\bibinfo {volume} {103}},\ \bibinfo
  {eid} {242404} (\bibinfo {year} {2013}),\
  http://dx.doi.org/10.1063/1.4839395}\BibitemShut {NoStop}%
\bibitem [{\citenamefont {Meyer}\ \emph {et~al.}(2014)\citenamefont {Meyer},
  \citenamefont {Althammer}, \citenamefont {Gepr{\"a}gs}, \citenamefont {Opel},
  \citenamefont {Gross},\ and\ \citenamefont {Goennenwein}}]{MeyerSMRAPL}%
  \BibitemOpen
  \bibfield  {author} {\bibinfo {author} {\bibfnamefont {S.}~\bibnamefont
  {Meyer}}, \bibinfo {author} {\bibfnamefont {M.}~\bibnamefont {Althammer}},
  \bibinfo {author} {\bibfnamefont {S.}~\bibnamefont {Gepr{\"a}gs}}, \bibinfo
  {author} {\bibfnamefont {M.}~\bibnamefont {Opel}}, \bibinfo {author}
  {\bibfnamefont {R.}~\bibnamefont {Gross}}, \ and\ \bibinfo {author}
  {\bibfnamefont {S.~T.~B.}\ \bibnamefont {Goennenwein}},\ }\href {\doibase
  10.1063/1.4885086} {\bibfield  {journal} {\bibinfo  {journal} {Appl. Phys.
  Lett.}\ }\textbf {\bibinfo {volume} {104}},\ \bibinfo {eid} {242411}
  (\bibinfo {year} {2014})}\BibitemShut {NoStop}%
\bibitem [{\citenamefont {Chen}\ \emph {et~al.}(2016)\citenamefont {Chen},
  \citenamefont {Takahashi}, \citenamefont {Nakayama}, \citenamefont
  {Althammer}, \citenamefont {Goennenwein}, \citenamefont {Saitoh},\ and\
  \citenamefont {Bauer}}]{SMR:review:Chen:JPCM:2016}%
  \BibitemOpen
  \bibfield  {author} {\bibinfo {author} {\bibfnamefont {Y.-T.}\ \bibnamefont
  {Chen}}, \bibinfo {author} {\bibfnamefont {S.}~\bibnamefont {Takahashi}},
  \bibinfo {author} {\bibfnamefont {H.}~\bibnamefont {Nakayama}}, \bibinfo
  {author} {\bibfnamefont {M.}~\bibnamefont {Althammer}}, \bibinfo {author}
  {\bibfnamefont {S.~T.~B.}\ \bibnamefont {Goennenwein}}, \bibinfo {author}
  {\bibfnamefont {E.}~\bibnamefont {Saitoh}}, \ and\ \bibinfo {author}
  {\bibfnamefont {G.~E.~W.}\ \bibnamefont {Bauer}},\ }\href {\doibase
  10.1038/ncomms10452} {\bibfield  {journal} {\bibinfo  {journal} {J. Phys.:
  Condens. Matter}\ }\textbf {\bibinfo {volume} {28}},\ \bibinfo {pages}
  {103004} (\bibinfo {year} {2016})}\BibitemShut {NoStop}%
\bibitem [{Note1()}]{Note1}%
  \BibitemOpen
  \bibinfo {note} {The orientation of the sublattice moments is crucial for the
  SMR in non-collinear magnets. In the collinear phase, all magnetic moments
  are aligned along the same axis. Describing the SMR in terms of either the
  total or the sublattice magnetization is then equivalent.}\BibitemShut
  {Stop}%
\bibitem [{\citenamefont {Zhang}\ and\ \citenamefont
  {Zhang}(2012{\natexlab{b}})}]{ZhangPRB}%
  \BibitemOpen
  \bibfield  {author} {\bibinfo {author} {\bibfnamefont {S.~S.-L.}\
  \bibnamefont {Zhang}}\ and\ \bibinfo {author} {\bibfnamefont
  {S.}~\bibnamefont {Zhang}},\ }\href {\doibase 10.1103/PhysRevB.86.214424}
  {\bibfield  {journal} {\bibinfo  {journal} {Phys. Rev. B}\ }\textbf {\bibinfo
  {volume} {86}},\ \bibinfo {pages} {214424} (\bibinfo {year}
  {2012}{\natexlab{b}})}\BibitemShut {NoStop}%
\bibitem [{\citenamefont {Calhoun}\ \emph {et~al.}(1957)\citenamefont
  {Calhoun}, \citenamefont {Overmeyer},\ and\ \citenamefont
  {Smith}}]{CalhounPR}%
  \BibitemOpen
  \bibfield  {author} {\bibinfo {author} {\bibfnamefont {B.~A.}\ \bibnamefont
  {Calhoun}}, \bibinfo {author} {\bibfnamefont {J.}~\bibnamefont {Overmeyer}},
  \ and\ \bibinfo {author} {\bibfnamefont {W.~V.}\ \bibnamefont {Smith}},\
  }\href {\doibase 10.1103/PhysRev.107.993} {\bibfield  {journal} {\bibinfo
  {journal} {Phys. Rev.}\ }\textbf {\bibinfo {volume} {107}},\ \bibinfo {pages}
  {993} (\bibinfo {year} {1957})}\BibitemShut {NoStop}%
\bibitem [{\citenamefont {Harris}(1963)}]{Harris_spinwave}%
  \BibitemOpen
  \bibfield  {author} {\bibinfo {author} {\bibfnamefont {A.~B.}\ \bibnamefont
  {Harris}},\ }\href {\doibase 10.1103/PhysRev.132.2398} {\bibfield  {journal}
  {\bibinfo  {journal} {Phys. Rev.}\ }\textbf {\bibinfo {volume} {132}},\
  \bibinfo {pages} {2398} (\bibinfo {year} {1963})}\BibitemShut {NoStop}%
\bibitem [{\citenamefont {Barker}\ and\ \citenamefont
  {Bauer}(2016)}]{BarkerYIG}%
  \BibitemOpen
  \bibfield  {author} {\bibinfo {author} {\bibfnamefont {J.}~\bibnamefont
  {Barker}}\ and\ \bibinfo {author} {\bibfnamefont {G.~E.~W.}\ \bibnamefont
  {Bauer}},\ }\href {\doibase 10.1103/PhysRevLett.117.217201} {\bibfield
  {journal} {\bibinfo  {journal} {Phys. Rev. Lett.}\ }\textbf {\bibinfo
  {volume} {117}},\ \bibinfo {pages} {217201} (\bibinfo {year}
  {2016})}\BibitemShut {NoStop}%
\bibitem [{\citenamefont {Kikkawa}\ \emph {et~al.}(2015)\citenamefont
  {Kikkawa}, \citenamefont {Uchida}, \citenamefont {Daimon}, \citenamefont
  {Qiu}, \citenamefont {Shiomi},\ and\ \citenamefont {Saitoh}}]{KikkawaSSE}%
  \BibitemOpen
  \bibfield  {author} {\bibinfo {author} {\bibfnamefont {T.}~\bibnamefont
  {Kikkawa}}, \bibinfo {author} {\bibfnamefont {K.}~\bibnamefont {Uchida}},
  \bibinfo {author} {\bibfnamefont {S.}~\bibnamefont {Daimon}}, \bibinfo
  {author} {\bibfnamefont {Z.}~\bibnamefont {Qiu}}, \bibinfo {author}
  {\bibfnamefont {Y.}~\bibnamefont {Shiomi}}, \ and\ \bibinfo {author}
  {\bibfnamefont {E.}~\bibnamefont {Saitoh}},\ }\href {\doibase
  10.1103/PhysRevB.92.064413} {\bibfield  {journal} {\bibinfo  {journal} {Phys.
  Rev. B}\ }\textbf {\bibinfo {volume} {92}},\ \bibinfo {pages} {064413}
  (\bibinfo {year} {2015})}\BibitemShut {NoStop}%
\bibitem [{\citenamefont {Boona}\ and\ \citenamefont
  {Heremans}(2014)}]{Boona_Heremans}%
  \BibitemOpen
  \bibfield  {author} {\bibinfo {author} {\bibfnamefont {S.~R.}\ \bibnamefont
  {Boona}}\ and\ \bibinfo {author} {\bibfnamefont {J.~P.}\ \bibnamefont
  {Heremans}},\ }\href {\doibase 10.1103/PhysRevB.90.064421} {\bibfield
  {journal} {\bibinfo  {journal} {Phys. Rev. B}\ }\textbf {\bibinfo {volume}
  {90}},\ \bibinfo {pages} {064421} (\bibinfo {year} {2014})}\BibitemShut
  {NoStop}%
\bibitem [{\citenamefont {Keffer}\ and\ \citenamefont
  {Kittel}(1952)}]{KefferKittelAFMR}%
  \BibitemOpen
  \bibfield  {author} {\bibinfo {author} {\bibfnamefont {F.}~\bibnamefont
  {Keffer}}\ and\ \bibinfo {author} {\bibfnamefont {C.}~\bibnamefont
  {Kittel}},\ }\href {\doibase 10.1103/PhysRev.85.329} {\bibfield  {journal}
  {\bibinfo  {journal} {Phys. Rev.}\ }\textbf {\bibinfo {volume} {85}},\
  \bibinfo {pages} {329} (\bibinfo {year} {1952})}\BibitemShut {NoStop}%
\bibitem [{\citenamefont {Kittel}(1959)}]{Kittel_FerriMR}%
  \BibitemOpen
  \bibfield  {author} {\bibinfo {author} {\bibfnamefont {C.}~\bibnamefont
  {Kittel}},\ }\href {\doibase 10.1103/PhysRev.115.1587} {\bibfield  {journal}
  {\bibinfo  {journal} {Phys. Rev.}\ }\textbf {\bibinfo {volume} {115}},\
  \bibinfo {pages} {1587} (\bibinfo {year} {1959})}\BibitemShut {NoStop}%
\end{thebibliography}
\end{document}